\def\hii{\ion{H}{2}}

\newcommand{\beq}{\begin{equation}}
\newcommand{\eeq}{\end{equation}}
\documentclass[11pt,preprint]{aastex}
\slugcomment{Submitted to the Astrophysical Journal}
\shorttitle{NEON AND OXYGEN IN M33}
\shortauthors{CROCKETT ET AL.}

\begin{document}
\title{Neon and Oxygen Abundances in M33}

\author{Nathan R. Crockett} 
\affil{Steward Observatory, University of Arizona, Tucson, AZ 85721,}
\affil{and National Optical Astronomy Observatories, 950 N. Cherry Avenue,
Tucson, AZ 85719}
\email{crockett@noao.edu}

\author{Donald R. Garnett} 
\affil{Steward Observatory, University of Arizona}
\affil{933 N. Cherry Avenue, Tucson, AZ 85721}
\email{dgarnett@as.arizona.edu}

\author{Philip Massey\altaffilmark{1}}
\affil{Lowell Observatory} 
\affil{1400 W. Mars Hill Road, Flagstaff, AZ 86001}
\email{massey@lowell.edu}

\and

\author{George Jacoby\altaffilmark{1}} 
\affil{WIYN Observatory}
\affil{950 N. Cherry Avenue, Tucson, AZ 85179}
\email{jacoby@noao.edu}
\altaffiltext{1}{Visiting Astronomer, Kitt Peak National Observatory,
National Optical Astronomy Observatories, which is operated by the
Association of Universities for Research in Astronomy, Inc. (AURA)
under cooperative agreement with the National Science Foundation.}

\begin{abstract}

We present new spectroscopic observations of 13 \hii\ regions in the Local
Group spiral galaxy M33. The regions observed range from 1 to 7 kpc in
distance from the nucleus. Of the 13 \hii\ regions observed, the [O~III]
$\lambda 4363$ line was detected in six regions. Electron temperatures
were thus able to be determined directly from the spectra using the [O~III]
$\lambda\lambda 4959,5007/\lambda 4363$ line ratio. Based on these temperature
measurements, oxygen and neon abundances and their radial gradients were
calculated. For neon, a gradient of $-0.016\pm0.017$ dex/kpc was computed,
which agrees with the Ne/H gradient derived previously from ISO spectra.
A gradient of $-0.012\pm0.011$ dex/kpc was computed for O/H, much shallower
than was derived in previous studies. The newly calculated O/H and Ne/H
gradients are in much better agreement with each other, as expected from
predictions of stellar nucleosynthesis.  We examine the correlation between
the WC/WN ratio and metallicity, and find that the new M33 abundances do not
impact the observed correlation significantly. We also identify two new
He II-emitting \hii\ regions in M33, the first to be discovered in a
spiral galaxy other than the Milky Way. In both cases the nebular He~II
emission is not associated with Wolf-Rayet stars. Therefore, caution is
warranted in interpreting the relationship between nebular He~II emission
and Wolf-Rayet stars when both are observed in the integrated spectrum of
an \hii\ region. 
\end{abstract}

\section{Introduction}

Spectroscopic observations of emission lines in \hii\ regions is still a primary
source of information of element abundances in the interstellar medium of spiral
and irregular galaxies. In the metal-rich inner disks of spirals, however,
abundances derived from such measurements have had considerable uncertainty.
The main reason for this has been the lack of accurate electron temperature
estimates, $T_{e}$ . The reason $T_{e}$ estimates have remained uncertain for
so long is largely due to the fact that the auroral lines on which accurate
temperature estimates rely (e.g. [O~III] $\lambda 4363$, [N~II] $\lambda 5755$,
and [S~III] $\lambda 6312$), are intrinsically weak and thus difficult to detect.
This difficulty is compounded in high metallicity \hii\ regions ($\gtrsim$ 0.5
(O/H)$_{\sun}$) because of increased cooling via IR fine structure lines, which
reduces the electron temperature and makes the temperature-sensitive diagnostic
line ratios even more difficult to measure. 

In the absence of direct temperature measurements, past studies of the composition
of \hii\ regions in spirals have made use of derived relations between O/H and
a strong emission line ratio, such as
$R_{23}$ = ([O II]$\lambda$3727 + [O~III]$\lambda\lambda$4959,5007)/H$\beta$,
with the relation often ``calibrated'' at the metal-rich end via predictions
of photoionization models (e.g., Edmunds \& Pagel 1984, Kewley \& Dopita 2002).
Unfortunately, models for metal-rich nebulae are rather sensitive to details
of input parameters (stellar fluxes, density, dust), and so various strong-line
calibrations have differed by factors of 2-3 or more, provoking considerable
debate. Thus, direct temperature measurements would be a valuable resource
for understanding the strong-line - O/H relationship.

The Local Group galaxy M33 is an excellent candidate for studying gas phase
element abundances in the interstellar medium (ISM). Because of its proximity
to our own galaxy, wealth of \hii\ regions \citep{boul74, cour87}, and relatively
low inclination, it has been the subject of much study over the past 65 years.
\citet{all42} obtained the first spectroscopy of \hii\ regions in M33 and 
found a radial gradient in the strength of [O~II] and [O~III] emission lines
across the galaxy, and suggested that it could be due to a gradient in 
excitation. \citet{smith75} was among the first to successfully measure [O~III]
$\lambda$4363 and thus to derive electron temperatures for \hii\ regions in
several spiral galaxies, including M33. This supported the hypothesis of
\citet{sea71} that a composition gradient, not a gradient in the excitation,
was responsible for the radial variation of oxygen forbidden line strengths.

Since then several more studies have been carried out concentrating on
determining a comprehensive abundance distribution for M33. \citet{kwit81}
used the [O~III] $\lambda\lambda 4959,5007/\lambda 4363$ and [O~II]
$\lambda 3727/\lambda 7325$ line ratios to compute electron temperatures
for seven \hii\ regions. They derived an oxygen gradient of --0.13 dex/kpc
and noted that Ne, N, S, and Ar followed essentially the same gradient.
\citet{vilc88} combined IPCS and CCD observations and derived electron
temperatures for [O~II], [O~III], and [S~III] for five giant \hii\ regions.
They derived a similar oxygen gradient of --0.12 dex/kpc. \citet{garn97}
compiled data collected by the above studies and recalculated abundances
using updated atomic coefficients, and obtained an oxygen abundance gradient
of $-0.11 \pm 0.02$ dex/kpc for M33.

Oxygen and neon are both produced mainly in stars larger than 10 solar 
masses, which have very short lifetimes. From stellar nucleosynthesis
calculations (e.g., Woosley \& Weaver 1995), it is expected that the
abundances of neon and oxygen should trace one another closely; the
chemical evolution calculations of Timmes, Woosley, \& Weaver (1995),
predict that the O/Ne ratio varies by no more than 0.1 dex over the
range $-$1.5 $<$ [Fe/H] $<$ 0.0 (which corresponds roughly to O/H in
the range 0.1 to 1.0 times the solar value). This prediction is 
supported by measurements of abundances in planetary nebulae (Henry 1990),
which show that Ne/O is constant over a wide range of O/H.

However, \citet{will02} derived neon abundances for 25 \hii\ regions in
M33 from infrared spectroscopy of [Ne~II] and [Ne~III] fine structure lines
using the $Infrared~Space~Observatory~(ISO)$, and reported a Ne/H gradient
($-0.034\pm0.015$ dex/kpc) that is significantly shallower than the O/H
gradient obtained from the visible-light measurements. The implication 
is that either set of measurements is in error, or that the Ne/O ratio 
across M33 varies by 0.5 dex over a factor ten in O/H. In this paper, we
report visible-light CCD spectroscopy of \hii\ regions in M33 to directly
measure electron temperatures, and if possible to resolve the discrepancy
between the results from infrared and visible-light spectra. 

\section{Observations and Data Reduction}

Spectra of 13 \hii\ regions were obtained on September 27-29, 1992 using the 
Mayall 4 m telescope at Kitt Peak National Observatory. The data were taken
using the R-C CCD spectrograph with a spectral range of 3600-5100 \AA\ at
approximately 2 \AA\ resolution with a slit width of 1\farcs8. The long slit
of the spectrograph was often oriented so that multiple \hii\ regions could
be observed at once. Risley prisms were used to compensate for not being at
the parallactic angle. A list of
the \hii\ regions observed for this study is given in Table 1. It presents
the coordinates for the center of each object, the position angle of the
slit in degrees measured from North through East, and the deprojected
distance of each \hii\ region from the nucleus. The coordinates were measured
from Mosaic H$\alpha$ imaging of M33 obtained by Massey et al. for the
Survey of Resolved Stellar Content of Local Group
Galaxies\footnote{http://archive.noao.edu/nsa/massey.html}.
We assumed a distance of 832 kpc to M33 \citep{wils90}, an inclination
angle of 56$\degr$ and a position angle of 23$\degr$ \citep{zar89}. The
nucleus of M33 was measured to be RA = 1$^{\rm h}$ 33$^{\rm m}$ 50.9$^{\rm s}$,
Dec = +30$^\circ$ 39$^\prime$ 37$\farcs$2 (J2000) from the Mosaic H$\alpha$
image.

The data reduction followed standard procedures. We divided a normalized
two-dimensional flat field, constructed using a combination of quartz lamp
exposures to derive the pixel-to-pixel sensitivity variations plus a
twilight sky exposure to correct for vignetting, into each spectrum.
Comparison of the normalized flat field frames taken at different times
during the run showed that the flat fields were reproducible to a level
of about 0.6\% r.m.s. The spectra were rectified and placed on a wavelength
scale using exposures of a He-Ne-Ar lamp; a cubic fit to the dispersion
data gave residuals of approximately 0.2 \AA\ r.m.s. Flux calibration of
the spectra was performed using observations of the spectrophotometric
standard stars Wolf 1346, BD+28 4200, Feige 110, G191B2B, Hiltner 600,
and Feige 34, all of which were observed at least once during the course
of each night. We derived the instrument sensitivity function by fitting a
low-order spline fit to the standard star measurements. We corrected all
of the standards and object spectra for atmospheric extinction using mean
KPNO extinction coefficients. Since the data were taken through a narrow
slit, absolute spectrophotometry was not possible, and we averaged the
various standard star sensitivity measurements at each wavelength to 
derive the sensitivity function. Based on the residuals of the data from
the fit, the relative calibration across the spectrum is good to about
2.5\% .

Emission line strengths were measured using the SPLOT task in IRAF's
ONEDSPEC package. The emission lines were fit with Gaussian profiles
and integrated in order to obtain a measurement of the flux. A linear
background was also fit and subtracted from the profile. In the instance
when two lines overlapped but were still resolved (e.g. the [O~III]
$\lambda 4959$ and $\lambda 5007$ lines), the line strengths were
deblended by fitting the profiles concurrently with the constraint
that both lines be fit with the same FWHM. 

Interstellar reddening toward each \hii\ region was determined by comparing 
observed hydrogen Balmer emission line ratios to the theoretical Balmer
decrement of Hummer and Storey (1987). The average interstellar extinction
curve, $f(\lambda)-f(H \beta)$, was taken from Savage and Mathis (1979). We
corrected the Balmer line strengths for underlying photospheric absorption
by incrementing the equivalent width of stellar absorption in 0.1 \AA\  
intervals from 0.1 \AA\ to 3.0 \AA. For each value of the equivalent width,
we determined the logarithmic reddening coefficient, $c(H \beta)$, for
the $I(H\gamma)/I(H\beta)$, $I(H\delta)/I(H\beta)$, and $I(H9)/I(H\beta)$
line ratios, giving us three independent estimates for $c(H \beta)$. The
adopted value for the equivalent width of underlying absorption was chosen
such that the dispersion in $c(H \beta)$ for each object was minimized.
The weighted mean of the three logarithmic reddening coefficients was then
computed and used to correct the line strengths for interstellar reddening.
The error in the weighted mean was taken as the uncertainty.

The dereddened emission line strengths for each object are listed in
Tables 2 and 3. The adopted logarithmic reddening coefficient and the
equivalent width of underlying photospheric absorption are also listed
for each object. The method by which we computed the uncertainties in
the dereddened line strengths and the logarithmic reddening coefficients
is described in Section 3.3.      

\section{Emission Line Analysis}

Electron temperatures, densities, and ionic abundances were computed using
the N-level atom code found in IRAF's NEBULAR package, a detailed description
of which can be found in \citet{shaw95}. This code is based upon the five-level
atom code of \citet{dero87}, but is more robust in that it allows for more
than five levels for most ions.

\subsection{Electron Temperatures and Densities}

Table 4 lists the derived measurements or upper limits for electron density,
$n_{e}$, estimated from the [O~II] $\lambda 3726/\lambda 3729$  line ratio. 
All $n_{e}$ calculations were performed using the IRAF task TEMDEN, 
assuming an electron temperature of 10,000 K. All but one of the observed
\hii\ regions (BCLMP691) had an [O~II] line ratio with a $1 \sigma$ uncertainty
which included the theoretical low density limit. As a result, we computed
$2\sigma$ upper limits for $n_{e}$ for all sources but BCLMP691 based on the
formal uncertainties in the dereddened [O~II] line ratios. The derived estimates
and upper limits for $n_{e}$ place all of the \hii\ regions in the low density
regime ($n_{e}\lesssim300$cm$^{-3}$). 

The commonly used [O~III] $\lambda\lambda 4959,5007/\lambda 4363$ line ratio 
was the only diagnostic available for deriving the electron temperature. Of 
the 13 spectra collected, six contained a detectable [O~III] $\lambda 4363$ 
line. We were thus able to compute electron temperatures for just under half
of the \hii\ regions observed. Electron temperatures were computed by again
using the IRAF task TEMDEN assuming an electron density of $n_{e}$ = 100
cm$^{-3}$. Table 4 lists the derived values for T[O~III] along with their
uncertainties.


\subsection{Ion and Element Abundances}

We estimated ion abundances for the six \hii\ regions which had a T[O~III]
measurement. Table 5 lists the ionic abundances which we were able to
derive: O$^{+}$, O$^{+2}$, and Ne$^{+2}$, along with their uncertainties.
All abundance calculations were performed using the IRAF task IONIC. 
Abundances for O$^{+}$ and O$^{+2}$ were computed using the summed line
ratios of [O~II] $\lambda\lambda 3726, 3729$ and [O~III] $\lambda\lambda
4959,5007$ respectively. Abundances for Ne$^{+2}$ were computed using the
[Ne~III] $\lambda 3869$ line. All of the regions studied have electron
densities too low for collisional de-excitation of the forbidden lines
to be important.

We assumed a two-zone model for the electron temperature structure for 
the abundance calculations. From ionization models, \citet{garn92} found
that the temperatures in the Ne$^{+2}$ and O$^{+2}$ zones were essentially
equal; therefore, we used T[O~III] to compute abundances for O$^{+2}$
and Ne$^{+2}$, while for O$^+$ we used estimates for T[O~II] based on
the relation between T(O$^+$) and T(O$^{+2}$) from \citet{camp86} and
\citet{garn92},
\begin{equation}
t(O^{+})=0.7t(O^{+2})+0.3,
\end{equation}
where $t=T_{e}/10^{4} K$.    
(In this format, temperatures specified by the 
ionic species, e.g. T(O$^{+2}$), refer to the ion-weighted average
temperature for that species, while temperatures specified by
spectroscopic notation, e.g. T[O~III], refer to temperatures 
computed directly from the spectra.)


Table 5 also lists the total element abundances for neon and oxygen
along with their uncertainties. We assumed that all of the oxygen
and neon was either in the singly or doubly ionized state, except
in the case of the two newly identified He~II emitting regions. 
For those objects, we assumed that O$^{+3}$/O $\approx$ He$^{+2}$/He,
based on the similarity in their ionization potentials. This should be
an accurate approximation when the He$^{+2}$ fraction is small, and we
indeed find that only a very small fraction (3\% for BCLMP090 and 6\%
for MA1) of oxygen is triply ionized in these regions. The ionization
fraction for Ne$^{+3}$ is even smaller because of its larger ionization
potential.

If we assume that the emission is optically thin, the total element
abundance is equal to the sum of the singly and doubly ionized components.
This computation is easily carried out for oxygen. Neon, however, presents
a problem because [Ne~II] is not seen in the visible spectrum. Total
element abundances for neon were therefore estimated by computing an
ionization correction using photoionization models from \citet{stas01}.
For the purposes of this study we did used a subset of their model
grid: we used the instantaneous burst models with Salpeter IMF and
upper stellar mass limit 120 solar masses (model sequences IKF, IKI,
and IKL). These model sequences were further restricted to those with
refractory elements depleted onto grains (depletion = 0.1 times the
normal values). 
Figure 1 is a plot of $X(Ne^{+2})/X(O^{+2})$ vs. $X(O^{+2})$ as predicted
by the photoionization models, where $X(O^{+2})=O^{+2}/(O^{+}+O^{+2})$ and
$X(Ne^{+2})=Ne^{+2}/(Ne^{+}+Ne^{+2})$. According to the models, for
$X(O^{+2})\gtrsim0.05$, the ratio $X(Ne^{+2})/X(O^{+2})$ is approximately
one. Therefore, we assume 
\begin{equation}
\frac{Ne}{H} = \left(\frac{O^{+}+O^{+2}}{O^{+2}}\right)\frac{Ne^{+2}}{H^+}.
\end{equation}

To add additional points and better constrain the neon and oxygen
abundance gradients, we included several \hii\ regions having electron
temperature measurements from \citet{vilc88}.
To be consistent in the atomic data used, we re-computed electron
temperatures and abundances for the \citet{vilc88} objects in the
same way as for our new sample. For all but two of the V\'\i lchez sources,
there was an [O~III] $\lambda\lambda 4959,5007/\lambda 4363$ line ratio
available for computing T[O~III]. For the two sources, NGC595 and MA2,
where T[O~III] could not be measured, the [S~III] $\lambda\lambda
9069,9532/\lambda 6312$ and [O~II] $\lambda 3727/\lambda 7325$ line
ratios were employed respectively to derive electron temperatures. For
MA2, we used the calibration,   
\begin{equation}
T(S^{+2}) = 0.83T(O^{+2}) + 1700 K
\end{equation}    
derived in \citet{garn92} to estimate $T(O^{+2})$. Table 6 lists the
measured electron temperature along with the deprojected distance from
the nucleus of each \hii\ region in the V\'\i lchez study from which an
electron temperature could be measured. Table 7 lists the ion and element
abundances for those sources.

\subsection{Uncertainties}

Uncertainties in the line fluxes, $\sigma_{line}$, were computed by adding
in quadrature the statistical uncertainty introduced by the photon noise,
$\sigma_{stat.}$, the uncertainty in the flat fielding correction,
$\sigma_{ff}$, and the uncertainty in the photometric calibration, $\sigma_{phot}$,
\begin{equation}
\sigma_{line}=\sqrt{\sigma_{stat}^{2} + \sigma_{ff}^{2} + \sigma_{phot}^{2}.}
\end{equation}
The statistical uncertainty was estimated from the noise in the continuum
using the formula (P\'erez-Montero \& D\'\i az 2003), 
\begin{equation}
\sigma_{stat}=\sigma_{cont}N^{1/2}\left(1 + \frac{EW}{N\delta\lambda}\right)^{1/2}
\end{equation}
where $\sigma_{cont}$ is the r.m.s. dispersion in the continuum, $N$ is the 
number of pixels in the window over which the line was measured, $EW$ is the
equivalent width of the line in \AA, and $\delta\lambda$ is the dispersion
in \AA/pixel. The first term in the sum represents the contribution of the
continuum noise to the total statistical uncertainty, while the second term
represents the contribution of the photon noise from the line itself. The
r.m.s. uncertainty in the flat field was approximately 0.6\%, determined by
comparing flat field exposures taken at different times during the run. The
relative end-to-end uncertainty in the flux calibration across the spectrum
was approximately 2.5\%. 

The error in $c(H \beta)$ was computed by formally propagating the uncertainty
from the line strengths, $I(\lambda)$,
\begin{equation}
\sigma_{c(H \beta)}=\sigma_{I_{\lambda}/I_{H\beta}}
\left[\frac{1}{ln(10)[f(\lambda)-f(H\beta)](I_{\lambda}/I_{H\beta})}\right]
\end{equation}
Once the errors in the interstellar reddening coefficients were computed, the 
uncertainties in the dereddened line ratios, $I_{\lambda_{o}}/I_{H\beta_{o}}$,
could be determined by again formally propagating the uncertainties,
\begin{equation}
\sigma_{I_{\lambda_{o}}/I_{H\beta_{o}}}=10^{c(f(\lambda)-f(H\beta))}
\sqrt{\sigma_{I_{\lambda}/I_{H\beta}}^{2}+
\sigma_{c}^{2}(I_{\lambda}/I_{H\beta})^{2}(f(\lambda)-f(H\beta))^{2}ln(10)^{2}}.
\end{equation}

We estimated the uncertainty in the electron temperature using a Monte
Carlo simulation. Assuming that the error in the [O~III] $\lambda\lambda
4959,5007/\lambda 4363$ line ratio can be accurately characterized by a
Gaussian profile, a measured line ratio and the error in that ratio can be 
interpreted respectively as the mean and standard deviation of a Gaussian
distribution. This distribution can then be populated by means of a Monte
Carlo simulation which generates deviates drawn from a Gaussian distribution.
The method used to compute the deviates is described in section 5.4 of
\citet{bev03}.

We populated each distribution with 10,000 deviates and fed them into the
N-level atom code to produce a corresponding distribution of electron
temperatures. Because of the logarithmic dependence of $T_{e}$ on its
diagnostic line ratio, the temperature distributions were asymmetric and
skewed such that the expectation value for a distribution was offset from
the value computed directly from the line ratio. The line ratios with the
largest uncertainties yielded the most highly skewed and biased temperature
distributions. For BCLMP706 this offset was most prominent ($\sim$ $-$500 K);
BCLMP745 and MA2 also had large offsets ($\sim$ $-$400 K). The rest of the
\hii\ regions had small temperature offsets ($<$ 200 K).  We therefore
chose to determine the expectation values for the temperature distributions
from the histogram of each distribution. Note that in the case of the
largest $T_e$ difference (BCLMP706), the change in O/H is less than
0.1 dex.

Because the temperature distributions were asymmetric, using a single
Gaussian to characterize the uncertainty in an electron temperature
measurement was not appropriate. We therefore used the expectation
value to divide the distribution into two sections. Deviates less than
the expectation value were used to characterize the negative error and
deviates greater than the expectation value were used to characterize
the positive error.
We assumed that the profile in each case could be modeled by one-half
of a Gaussian profile, and used a fit to determine the one-sided 
standard deviation for the distribution on either side of the expectation
value. The two values for the standard deviation thus derived were 
taken to be the negative and positive uncertainties for $T_e$.


The uncertainty in the ionic abundances were computed using the same technique.
The total error in these measurements, however, came from two sources: 
the error in the emission line flux (which we assume is normally distributed)
and the error in the electron temperature (which has a skewed distribution
as described above). We therefore modeled the abundance errors with a 
Monte Carlo analysis, feeding in both sources of error. The resulting
uncertainties are those listed in Tables 5 and 7.

\section{Two New He II-emitting \hii\ Regions in M33}

We report here the discovery of two new examples of \hii\ regions
that emit nebular He~II $\lambda$4686 emission in M33. Nebular
He~II emission is a phenomenon seen in only a handful of \hii\
regions in Local Group galaxies (Garnett et al. 1991). No single
cause has been identified for the source of the He~II emission
in such objects; He~II emission has been observed in nebulae
associated with WO and peculiar WN stars, high-mass X-ray
binaries, and in one case with an otherwise ordinary mid-O type
main sequence star (Pakull \& Angebault 1986; Stasi\'nska et al.
1986; Garnett et al. 1991; Garnett et al. 2000). To date, these
He~II nebulae have been found
primarily in metal-poor dwarf irregular galaxies, except for
the Galactic Wolf-Rayet bubble G2.4+1.4 associated with the
WO star WR 102 (Dopita et al. 1990). Narrow He~II emission
has also been found in a number of metal-poor dwarf starburst
galaxies, as catalogued in Schaerer, Contini, \& Pindao (1999).
The two objects discovered in this study are the first He~II
nebulae identified in a spiral galaxy beyond the Milky Way.

The first new He~II region in M33 is BCLMP090, a small, relatively
faint \hii\ region about 1 kpc from the center of the galaxy.
BCLMP090 has a diameter of roughly 7$\arcsec$ as measured from 
the H$\alpha$ image from the MOSAIC Survey of Local Group Galaxies.
The nebular morphology consists of a knot on the northwest side,
opening out into a ring toward the southeast. Its diameter of
about 30 pc clearly shows that this is not a planetary nebula.
Figure 2 shows a portion of our spectrum of BCLMP090. The plot
shows the detections of He~II $\lambda$4686 and [O~III] $\lambda$4363;
[Ar~IV] $\lambda$4740 may also be detected marginally, which would
be consistent with the presence of He~II emission. The high
[O~III] $\lambda$5007/H$\beta$ ratio of 6.6 (see Table 2) is very
uncharacteristic of \hii\ regions with similar O/H, providing
further support for the high excitation and ionization state of
BCLMP090. 

Because of the narrow spectrograph slit used, we are not able to
measure total $\lambda$4686 and H$\beta$ fluxes very accurately.
The fluxes measured within the slit, corrected for reddening, are
F($\lambda$4686) =
3.1$\times$10$^{-16}$ erg cm$^{-2}$ s$^{-1}$
and F(H$\beta$) = 
1.1$\times$10$^{-14}$ erg cm$^{-2}$ s$^{-1}$.
Examination of the spatial profile across the slit showed that the
He~II emission covered a region about 2\farcs8 across (FWHM). 
With a 1\farcs8 slit, this indicates that the total $\lambda$4686
flux is probably no more than two times larger than the measured
flux within the slit. H$\beta$ is emitted over a much larger area,
and we crudely estimate that the total F(H$\beta$) is 3-4 times
larger than the value measured within the slit. Assuming a distance
of 832 kpc to M33, we thus estimate L($\lambda$4686) $\approx$
5$\times$10$^{34}$ erg s$^{-1}$
and L(H$\beta$) $\approx$
3$\times$10$^{36}$ erg s$^{-1}$.
These values are slightly lower than, but consistent with, those
measured for other He~II nebulae, while the L($\lambda$4686)/L(H$\beta$)
ratio is entirely consistent with those for other objects (Garnett
et al. 1991).

The observed continuum in BCLMP090 is faint and shows no obvious
stellar features other than Balmer absorption lines, so we are 
not able to provide a classification for the star, except that it
does not appear to be a Wolf-Rayet star. The measured L(H$\beta$)
is consistent with that of a late O-type main sequence star. We
searched the catalogs of X-ray sources in M33 derived by Pietsch
et al.  (2004) from $XMM-Newton$ observations, and from $Chandra$
observations by Grimm et al. (2005). We found no source within
one arcmin of the position of BCLMP090 with $L_X$ $>$ 
2$\times$10$^{34}$ erg s$^{-1}$.
This indicates that BCLMP090 is not an X-ray ionized nebula powered
by a currently-active X-ray binary, as exemplified by LMC-N159F,
which is powered by the massive binary LMC X-1 (Pakull \& Angebault
1986). 

The second He~II nebula we identified is MA 1, the outermost known 
\hii\ region in M33. MA 1 was observed previously by McCall, Rybski,
\& Shields (1985) and Garnett, Odewahn, \& Skillman (1992), who
demonstrated the low O/H of the nebula. MA 1 is a large \hii\ 
complex approximately 45$\arcsec$ in diameter, consisting of a
bright knot surrounded by extensive nebular filaments. Garnett 
et al. (1992) noted the presence of He~II $\lambda$4686 emission
in MA 1, but they were not able to determine definitively whether
the emission was nebular or stellar. 

With our new high signal/noise spectrum of MA~1, we are able to
confirm that the narrow He~II emission is spatially extended, and
thus is nebular in nature. Figure 3 shows spectra from two 
positions in MA~1. The first position corresponds to the bright
knot and includes the stellar object ``star 1'' noted by Garnett
et al. (1992), which is probably a star cluster. At this position
the spectrum shows the nebular He~II emission plus a strong stellar
continuum with a number of absorption features (e.g., He~II
$\lambda$4512, Si~IV $\lambda\lambda$4089,4116, He~I $\lambda\lambda$
4120,4143,4387,4471) characteristic of OB stars. The second spectrum,
from a position 9$\arcsec$ E of the first shows no nebular He~II
emission, but rather the broad He~II feature characteristic of a
WN star, one of two spectroscopically confirmed WN stars (MC 8 and
MC 9) in MA~1 (Massey \& Conti 1983).

Meanwhile, Figure 4(a) shows the spatial profile of both the $\lambda$4686
emission and the stellar continuum, while Figure 4(b) shows the spatial
profile of the He~II emission with the stellar continuum subtracted -- 
that is, only pure nebular He~II emission or the W-R $\lambda$4686
feature. The regions of nebular He~II emission and the position of
the WN star are marked. It is notable from Figures 3 and 4 that the
nebular He~II emission is not strongest at the position of the W-R
star, but rather at the position of the OB cluster, more than 30 pc
away. While it is not possible to completely rule out that the W-R
star is powering the He~II emission, the poor spatial correlation
argues strongly against the W-R star being the source of He$^{+}$-ionizing
photons. We conclude that the He~II nebula is powered by a source(s)
within the OB star cluster. This distinction is significant, as narrow
(presumed nebular) He~II emission is often associated with the presence
of Wolf-Rayet stars (e.g., Izotov et al. 1997), and it is often argued
that this demonstrates that Wolf-Rayet stars are responsible for the
nebular emission. Our new observations of MA~1 provide evidence that
it is not safe to automatically associate nebular He~II emission with
Wolf-Rayet stars based on an integrated spectrum.

As we did for BCLMP090, we estimate roughly the He~II and H$\beta$
fluxes for MA~1, keeping in mind that this estimate is more uncertain
because the spectrograph slit subtends a much smaller fraction of MA~1.
We measured F($\lambda$4686) = 
1.2$\times$10$^{-15}$ erg cm$^{-2}$ s$^{-1}$
over a region about 8$\arcsec$ across. If we assume uniform surface
brightness over a circular region with diameter 8$\arcsec$, we derive 
L($\lambda$4686) $\approx$ 
5$\times$10$^{35}$ erg s$^{-1}$.
This value may be accurate to no more than a factor of two; nevertheless,
the derived flux is still similar to the values derived by Garnett et al.
(1991) for other He~II nebulae. A more precise value would require a
measurement of the He~II flux over the entire emission region. Such
measurements are desirable to accurately determine the He~II luminosities
and to constrain the parameters of the ionizing stars. 

Given the lack of association between the He~II emission in MA~1 and
the W-R star, it is not possible at present to identify the likely
source powering the He~II emission. The $XMM-Newton$ observations of
M33 show no X-ray point source at the location of MA~1 (Pietsch et al.
2004); the nearest point source is 13$\arcsec$ north of the region
we observed, making an association unlikely. (Grimm et al. 2005 did
not observe this field with $Chandra$.) More extensive observations
of these objects are desirable to identify the likely ionizing stars,
constrain their atmospheric parameters, and ultimately to produce a
much clearer picture of how He~II emission is produced in these objects
and whether or not these peculiar ionizing stars are natural products of
stellar evolution. 

\section{Oxygen and Neon Abundance Gradients}

The abundance measurements vs. deprojected galactocentric radius for both
O/H (top) and Ne/H (bottom) are displayed in Figure 5. We also show in
Figure 5 the data for \hii\ regions with measured electron temperatures
from V\'\i lchez et al. (1988). These two data sets give O/H gradient
slopes ($-$0.03$\pm$0.01 for our data vs. $-$0.004$\pm$0.03 for the
V\'\i lchez et al. points) and zeropoints ($-$3.62$\pm$0.05 vs.
$-$3.55$\pm$0.13) that agree to within 1$\sigma$; similarly, we
find agreement to within 1.3$\sigma$ for the Ne/H gradients. 
Therefore, although there appears to be a slight systematic offset 
between the two data sets, the offset is statistically insignificant,
and we combine the two sets of points to estimate the O/H and Ne/H
gradients for this and all subsequent discussion in this paper. The
dashed lines in Figure 5 represent the resulting O and Ne abundance
gradients based on a weighted linear least squares fit to the combined
data from this study and the five points from V\'\i lchez et al. (1988).
The least squares fit for the O/H gradient is $-0.012\pm0.011$ dex/kpc
and the least squares fit for the Ne/H gradient is $-0.016\pm0.017$
dex/kpc. The newly-derived O/H and Ne/H gradients are in good agreement
with each other (within 1 $\sigma$).

\citet{will02} used the [Ne~II] 12.81 $\mu$m and [Ne~III] 15.56 $\mu$m
fine structure lines with radio continuum measurements to measure Ne
abundances in M33. These measurements have the advantage of being almost
completely temperature independent because both the fine structure
lines and the thermal continuum have similar temperature dependences.
From their sample of 25 \hii\ regions \citet{will02} derived a Ne/H
gradient of $-0.034\pm0.015$ dex/kpc. Our new measurement of the Ne
and O gradients are now only slightly more than 1$\sigma$ from the
\citet{will02} results. 

Our new oxygen gradient is much shallower than previous studies have indicated.
\citet{garn97} recompiled data from several studies in the literature 
\citep{smith75, kwit81, vilc88, gos92} which measured oxygen abundances
in \hii\ regions. They computed an oxygen gradient of $-0.11 \pm 0.02$
dex/kpc for M33. We note that many of the \hii\ regions in these studies
did not have measured electron temperatures, necessitating the use of 
a calibration of the strong [O~II] and [O~III] lines vs. O/H to estimate
abundances (Edmunds \& Pagel 1984). Several recent studies (Castellanos
et al. 2002; Kennicutt, Bresolin, \& Garnett 2003; Bresolin, Garnett,
\& Kennicutt 2004) have now derived $T_e$, from measurements of [N~II]
and [S~III] line ratios, in a number of metal-rich \hii\ regions in
NGC 1232, M101, and M51. These studies have found \hii\ regions with
O/H as high as the solar value or slightly higher. They have also shown
that abundances based on directly measured electron temperatures are
significantly smaller (0.3-0.5 dex) than those derived from various
strong-line O/H calibrations. Our new results for \hii\ regions in M33
are consistent with these other studies, implying that the inner disk
in M33 is not as metal-rich as previously thought. We should emphasize,
however, that we have measured abundances for only one object within
2 kpc of the nucleus of M33. 



In Figure 6 we display both the neon abundances derived in the present
study and those derived by Willner \& Nelson-Patel (2002). The solid and
dashed lines are the weighted least squares fits to the Willner \&
Nelson-Patel data and our data, respectively. The two fits show an
offset in Ne/H, such that the Willner \& Nelson-Patel data are
approximately 0.2 dex higher than the results from the visible spectra;
this difference is significant at more than the 3$\sigma$ level.
If we remove BCLMP745, the \hii\ region with the lowest neon abundance
level, this result is largely unaffected because of the relatively
large uncertainty associated with this measurement. We disregard 
the difference in slope of the neon gradients because of the large
error in the value of the slope.

What could be the source of this difference? One possibility that we
considered was that the ionization corrections that we used to derive
the neon abundances were inaccurate, as they were estimated from 
photoionization models; the fractional ionization of Ne$^{+2}$ is 
very sensitive to the stellar radiation field beyond 40 eV, which
is still relatively uncertain for O star model atmospheres. If the
ionization corrections we used were systematically too small, then
we might expect to see an offset in $X(Ne^{+2})$ between our data
set and those measured by Willner \& Nelson-Patel (2002), since our
\hii\ regions should have similar ionization on average. Figure 7 plots
$X(Ne^{+2})$ as a function of galactocentric radius for our data
set and those from Willner \& Nelson-Patel (2002). The figure shows
that there is no systematic difference in $X(Ne^{+2})$ between the
two studies. The one exception is BCLMP090, which is a He~II nebula
and thus has a higher Ne ionization than expected for \hii\ regions
ionized by normal O and B stars. The results in Figure 7 suggest that
ionization corrections are probably not the source of the offset in
neon abundance.

Another possibility is that the electron temperature measurements
for this study were biased towards higher values, leading to an
underestimation of the metallicity. \citet{stas05} has shown that in
photoionization models for high metallicity \hii\ regions (log(O/H)
$\gtrsim$ -3.4), oxygen abundances based on direct measurements of
electron temperatures derived from [O~III] line ratios could lead
to biased oxygen abundance estimates. This is brought about by
temperature gradients within a nebula that are the result of strong
cooling by [O~III] and [N~III] fine-structure lines. The effect of
the bias is that abundances can be underestimated by as much as
0.2 dex when the true O/H is approximately solar (log O/H = --3.3).
However, this bias is very small at lower O/H, and so would not
explain the difference we see between the IR and visible-light
measurements at low O/H. A related possibility is that of temperature
fluctuations (Peimbert 1967), wherein fluctuations in $T_e$ lead
to a derived temperature that is too high (and thus abundances
that are too low) compared to the average temperature. Whether
temperature fluctuations large enough to affect abundances are
present in photoionized nebulae is still a subject of debate (see
e.g., Mathis 1995, Liu 2002).


While the discrepancy we see in Ne/H is of concern, we should
emphasize that the optical data comprise only a small set, and
we still do not have enough data points in the inner disk of M33 to
constrain the Ne and O gradients particularly well. A larger 
data set should settle the nature of the discrepancy. A program
is under way to collect new optical spectroscopy from the ground
and IR spectroscopy with the {\it Spitzer Space Telescope} to
evaluate the discrepancy we see here.

Figure 8 displays the radial distribution of O/H from all of the
\hii\ regions discussed in this paper ({\it filled squares}), along
with O/H measurements from spectroscopy of B supergiants from
Monteverde, Herrero, \& Lennon (2001; {\it stars}). Monteverde et
al. noted that the supergiant abundances were in reasonable agreement
with the results of V\'\i lchez et al. (1988). Nevertheless, from
examination of Figure 8 we can say that the radial trend in O/H
based on the supergiants is not discordant with what we find, either.
This is undoubtedly a result of the relatively large uncertainties
in the oxygen abundances. Many more spectroscopic measurements of
stars and \hii\ regions in M33 are needed to properly compare the
independent results from stellar and gas abundance studies.

A number of important astrophysical relations (such as the Cepheid
period-luminosity relation and the I(CO) - N(H$_2$) relation) have,
or may have, a significant dependence on metallicity. The metallicity
dependence in many cases is derived using abundances derived from 
\hii\ regions. Kennicutt, Bresolin, \& Garnett (2003) noted that the
derived metallicity dependence for such relations would steepen 
considerably if metallicity gradients in spiral galaxies are flatter
than usually supposed, as found in their paper, and in the present
work. Here we revisit briefly the correlation between Wolf-Rayet
star properties and metallicity.

It has long been known that the relative number of WC-type and
WN-type Wolf-Rayet (WR) stars varies with galactrocentric distance
within M33, with the center richer in those of type WC (Massey \&
Conti 1983, Massey \& Johnson 1998). This has been taken to be a
demonstration of the importance of stellar winds on the evolution
of massive stars; stars with higher initial oxygen abundances will
have a higher mass-loss rate, and it will hence be easier to produce
WC stars (Massey \& Johnson 1998). The trend observed within M33
also extends to such low metallicity galaxies such as the SMC, and
to the metal-rich galaxy M31 (Massey 2003).  A similar trend is
seen with metallicity in the relative number of WRs and red
supergiants (Massey 2003).  Modern stellar evolutionary theory
does a good job of predicting the trend, at least for the relative
number of WCs and WNs (Massey 2003, Meynet \& Maeder 2005). 

Figure 9 shows an updated version of Figure 11 from Massey (2003),
showing the relationship between the WC/WN number ratio and the
mean O/H. The filled circles represent averages over galaxies or
regions within the galaxies; see Massey (2003) and Massey \& Johnson
(1998) for a complete description of the data for this figure. The
oxygen abundances for several galaxies in this plot have been updated
with new values from the following sources: LMC and SMC (Garnett 1999);
M33 (this work); Milky Way (Deharveng et al. 2000). With the exception
of the data point for M31, all of the abundances for the galaxies in
Figure 9 have been derived directly from measurements of the electron
temperature; the three points for M33 refer to O/H values determined
from the least squares fit at 0.8 kpc, 2 kpc, and 5 kpc from the 
nucleus. The M31 metallicity, in contrast, was derived from the strong
[O~II] and [O~III] emission lines using a relation that is calibrated
largely with photoionization models.

We also plot in Figure 9 the recent results for the metallicity
dependence of the WC/WN ratio from Meynet \& Maeder (2005), who
computed the effects of rotation on mixing and stellar evolution.
The predicted relationship between WC/WN and O/H is depicted by
the region bounded by the dotted line (see also Figure 11 of 
Meynet \& Maeder 2005). The bounds of the region reflect the fact
that stars likely exhibit a range of rotation speeds; the upper
bound of the region represents main sequence stars rotating at
300 km s$^{-1}$, while the lower bound represents non-rotating stars.
We see that our new abundance values for M33 are still consistent
with the predicted correlation (keeping in mind that the WC/WN ratio
for the Galaxy is likely to be overestimated because of incompleteness
for WN stars -- see Massey 2003). While this is reassuring, we note
that the M31 data has a much greater impact on the slope of the
correlation. The fact that the M31 abundances were derived with a
different method raises some concern; as we have noted above, recent
measurements of electron temperatures for metal-rich \hii\ regions
yield significantly smaller abundances than strong-line calibrations.
Furthermore, analyses by Venn et al. (2000) and Smartt et al. (2001)
of four supergiants in M31 does suggest an oxgyen abundance similar
to that of the solar neighborhood, approximately 0.3 dex lower. We
have illustated this by the unfilled circle for M31 in Figure 9, in
which case the correlation between WC/WN and O/H would be considerably
steeper. Note, however, that the WC/WN ratio for the Milky Way is likely
too large due to selection effects (Massey \& Johnson 1998). A new,
deep spectroscopic study of \hii\ regions (and young massive stars)
in M31 would be worthwhile to evaluate this question.


\acknowledgments
PM and GJ thank Jennifer Hedden for her assistance at the telescope.
DRG thanks Katia Cunha for helpful discussions of abundances in blue
supergiants, and Steve Willner for discussions of the ISO measurements
for M33 \hii\ regions.

\clearpage

\begin{figure}
\epsscale{0.95}
\plotone{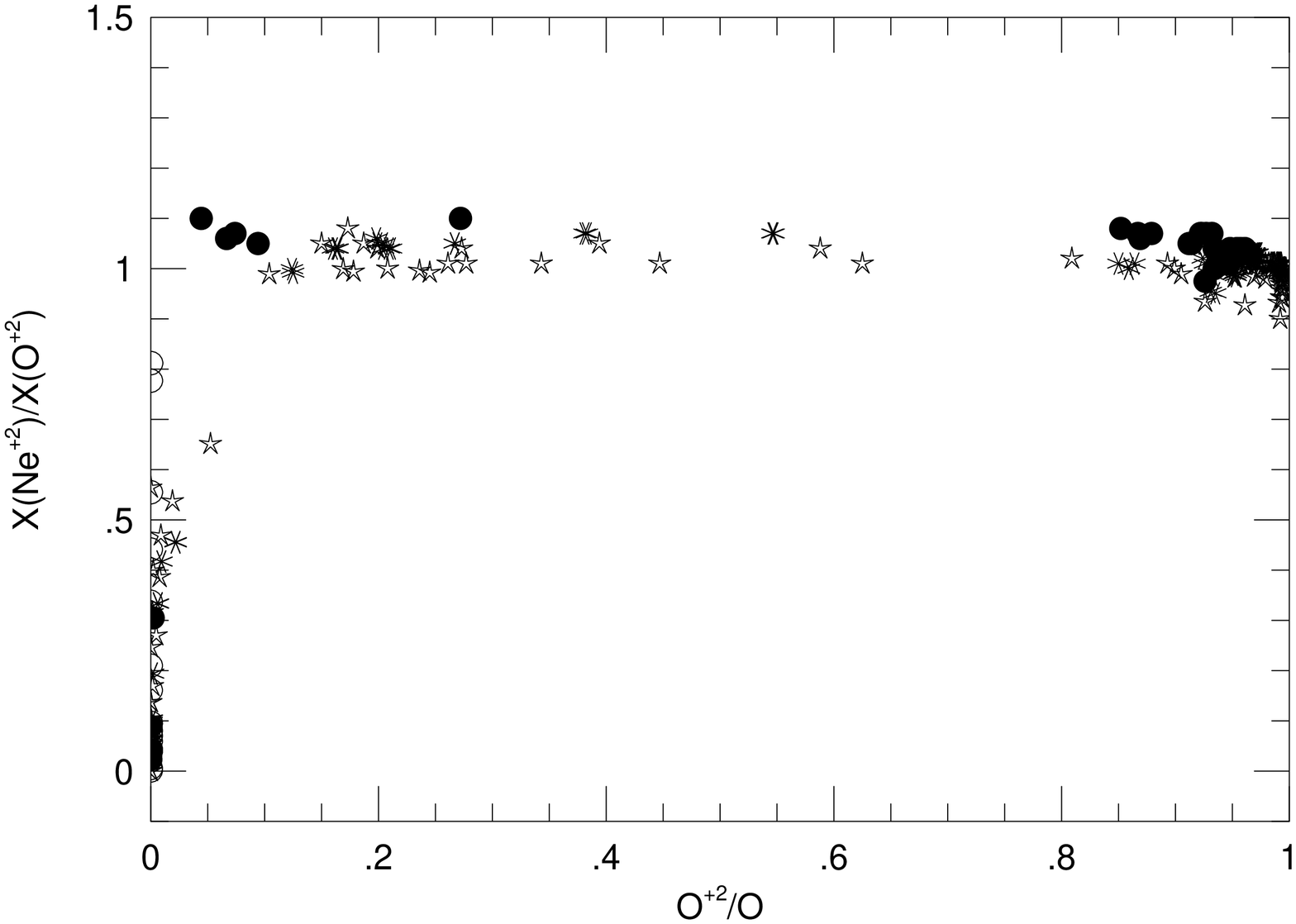}
\caption[xNe3xO3.ps]{The ion ratio $X(Ne^{+2})/X(O^{+2})$ plotted as a
function of oxygen ionization, $X(O^{+2})$, as computed by the models of
Stasi\'nska, Schaerer, \& Leitherer(2001). Filled circles represent 
the IKF model sequence, asterisks the IKI sequence, and stars the IKL
sequence, as discussed in the text. Based on this diagram, when
$X(O^{+2})\gtrsim0.05$, we can assume that $X(Ne^{+2}) \approx X(O^{+2})$.}
\end{figure}

\clearpage

\begin{figure}
\epsscale{0.9}
\plotone{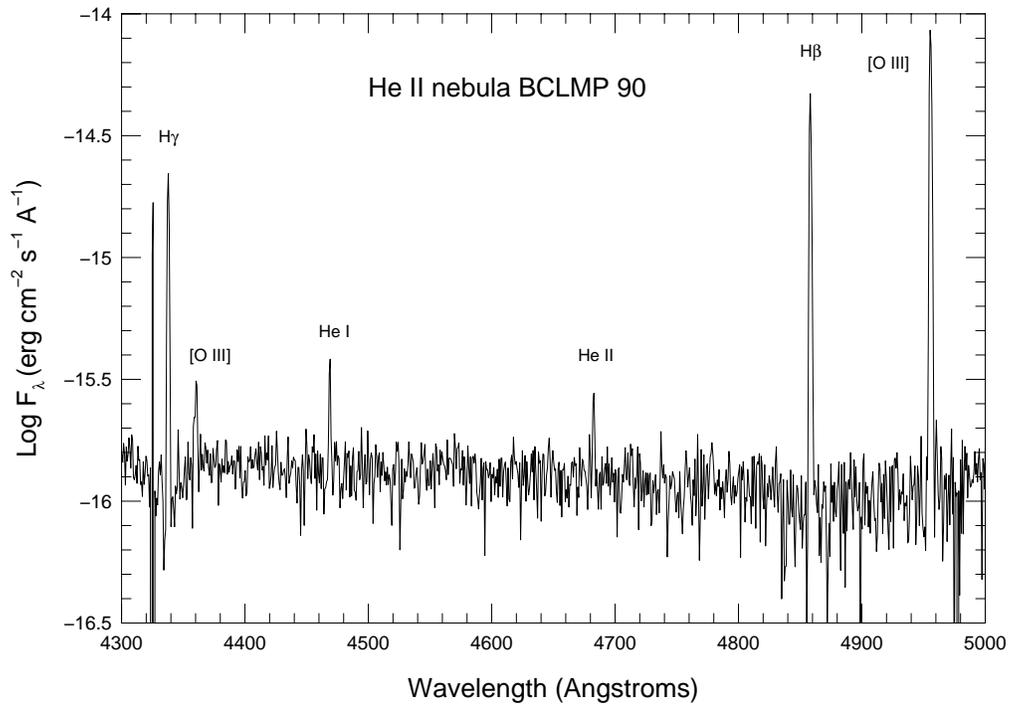}
\caption{A portion of our KPNO 4-m spectrum of the He~II nebula
BCLMP090 in M33. This segment has been expanded to show our detections
of He~II $\lambda$4686 and [O~III] $\lambda$4363 in this object. 
}
\end{figure}

\clearpage

\begin{figure}
\epsscale{0.9}
\plotone{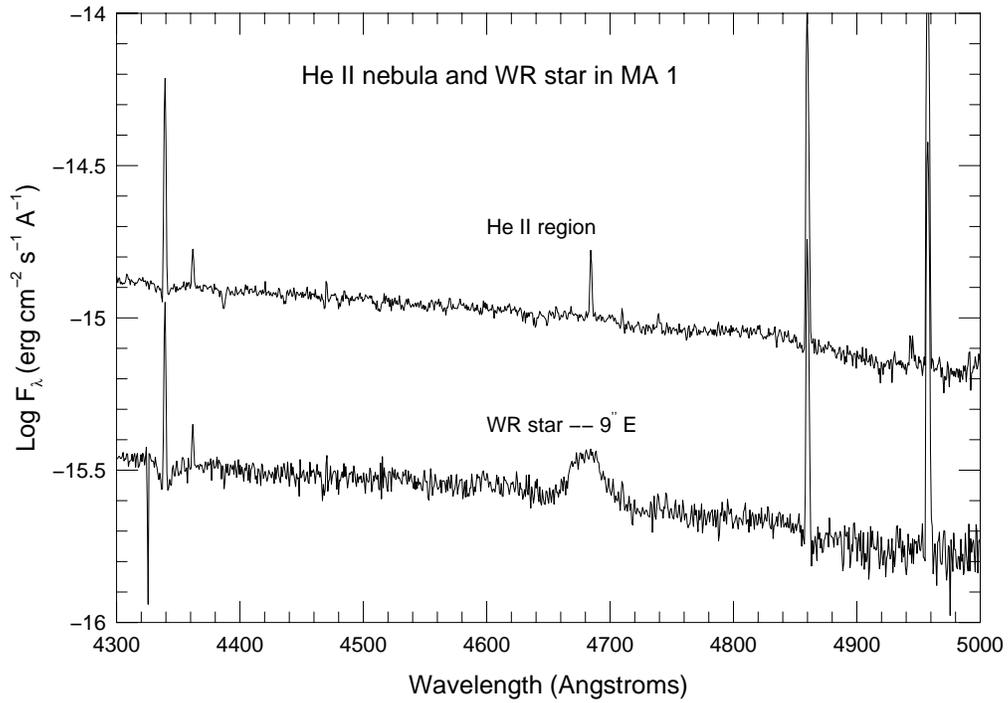}
\caption{A portion of our KPNO 4-m spectrum of the \hii\ region 
MA~1 in M33. The upper spectrum shows the spectrum of a region 
within MA~1 that encompasses the He~II region and the brighter
stellar OB cluster. The lower spectrum shows a part of MA~1 that
is 9$\arcsec$ west of the first position, which includes the 
WN star discussed in the text. Note the lack of narrow He~II
$\lambda$4686 line emission at the position of the WN star.
}
\end{figure}

\clearpage

\begin{figure}
\epsscale{0.9}
\plotone{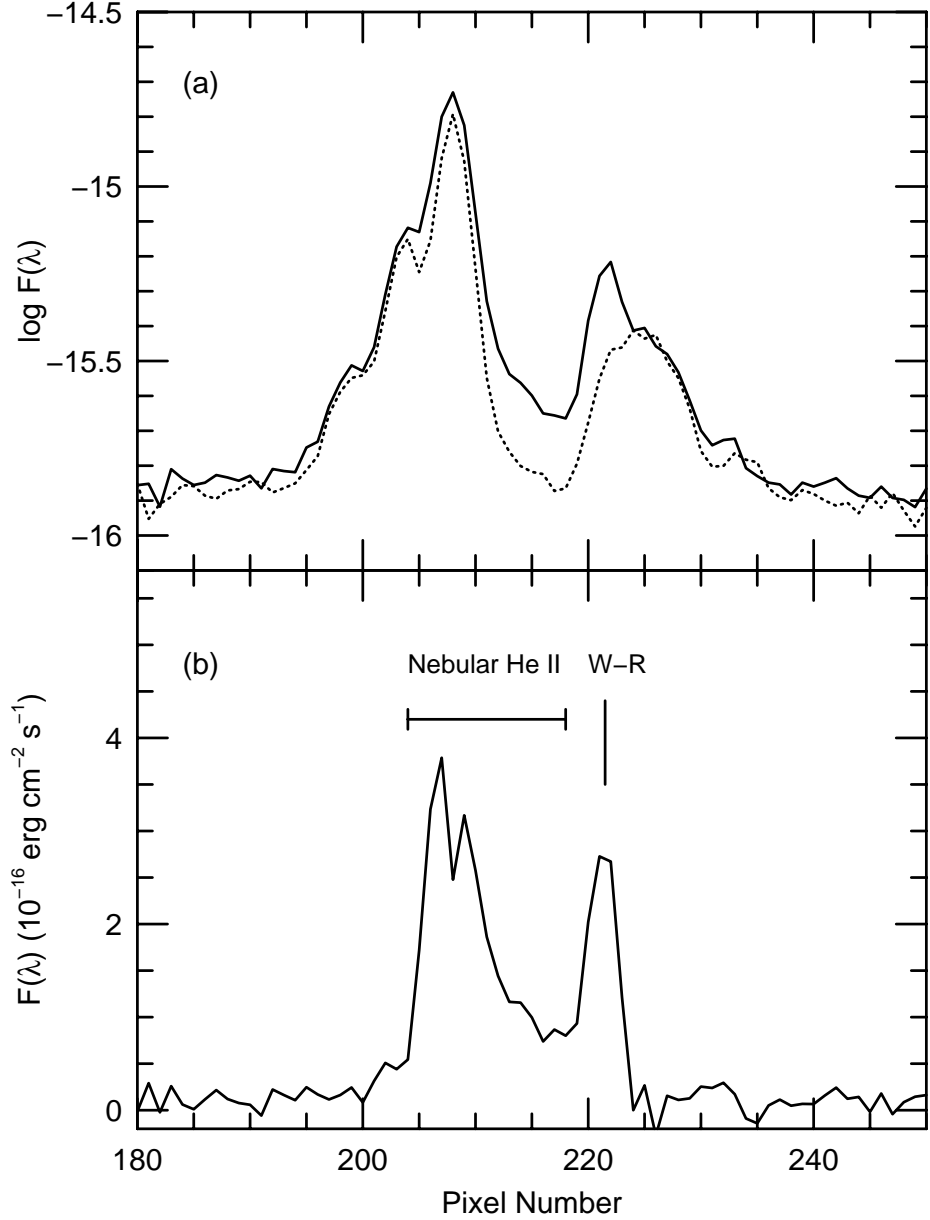}
\caption{(a) A spatial cut of our spectrum of MA~1 along the slit
showing the spatial variation of $\lambda$4686 ({\it solid line})
and stellar continuum ({\it dotted line}). (b) The spatial variation
of continuum-subtracted He~II $\lambda$4686 emission across MA~1.
The locations of the He~II nebular emission and the WN star are 
marked. The scale is 0.72 arcsec pixel$^{-1}$, and east is to the
left in the plots. 
}
\end{figure}

\clearpage

\begin{figure}
\plotone{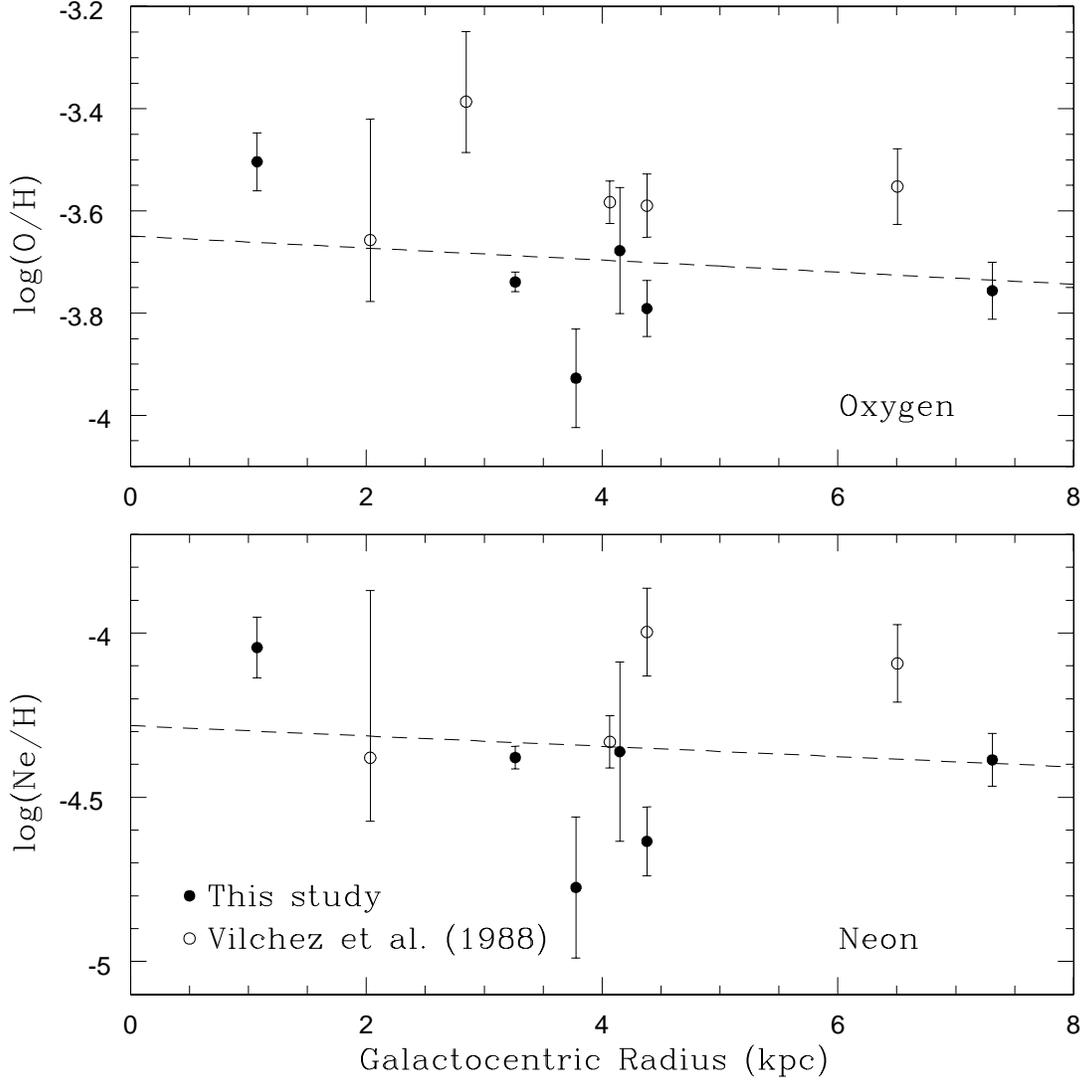}
\caption[bothgrad.ps]{log(O/H) (top) and log(Ne/H) (bottom) plotted as a
function of galactocentric radius. Filled circles are abundance measurements
from this study and open circles are measurements from V\'\i lchez et al.
(1988). The dashed lines in each panel represent the fitted abundance
gradients for oxygen and neon based on the combined data from both sets
of data plotted.
}
\end{figure}

\clearpage

\begin{figure}
\plotone{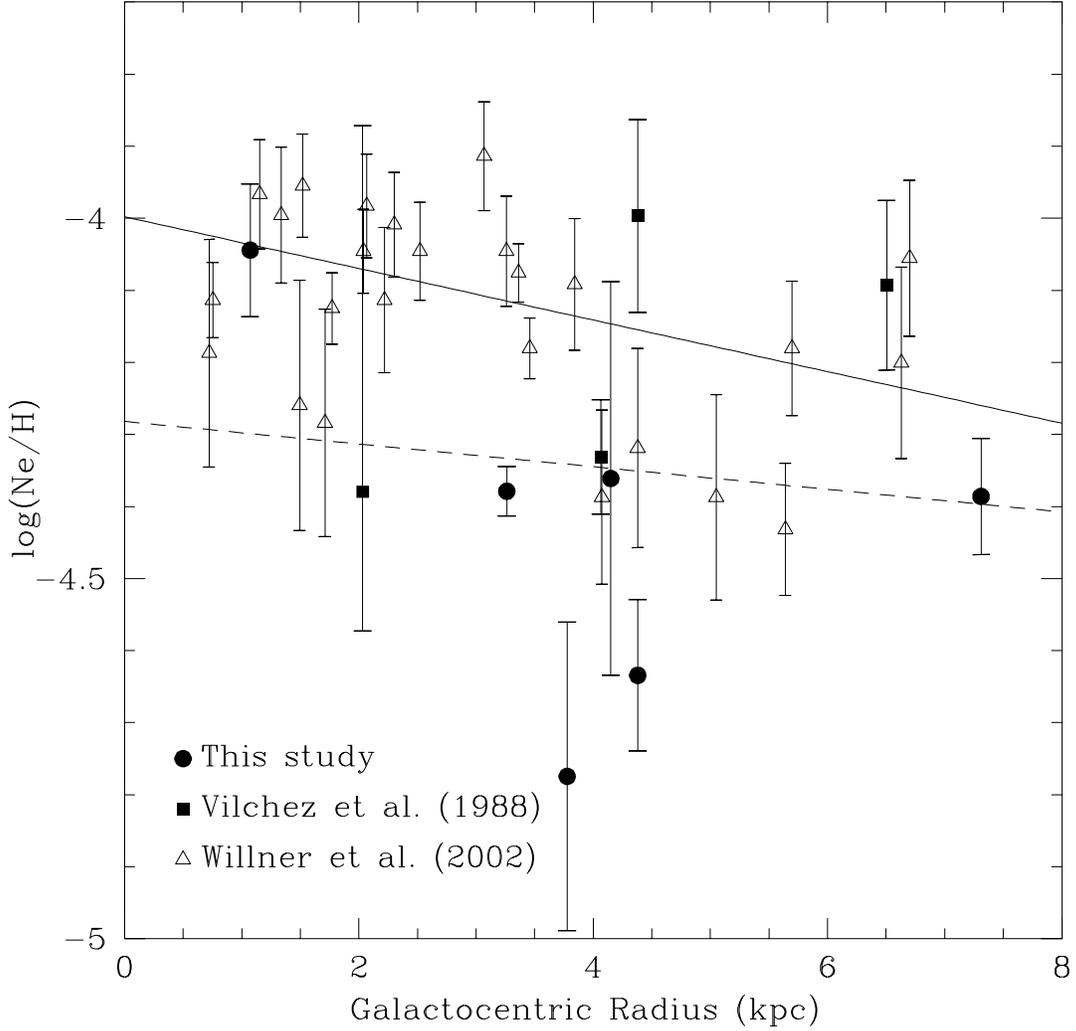}
\caption[wnp.ps]{Comparison of the overall neon abundances derived
by Willner and Nelson-Patel (2002) (triangles) to those derived in this
study and V\'\i lchez et al. (1988) (filled circles and filled squares,
respectively). The neon gradient derived from the Willner data is shown
as a solid line and the gradient derived from this study and the
V\'\i lchez data is shown as a dashed line. Note the offset between
the two gradients.}
\end{figure}

\clearpage

\begin{figure}
\plotone{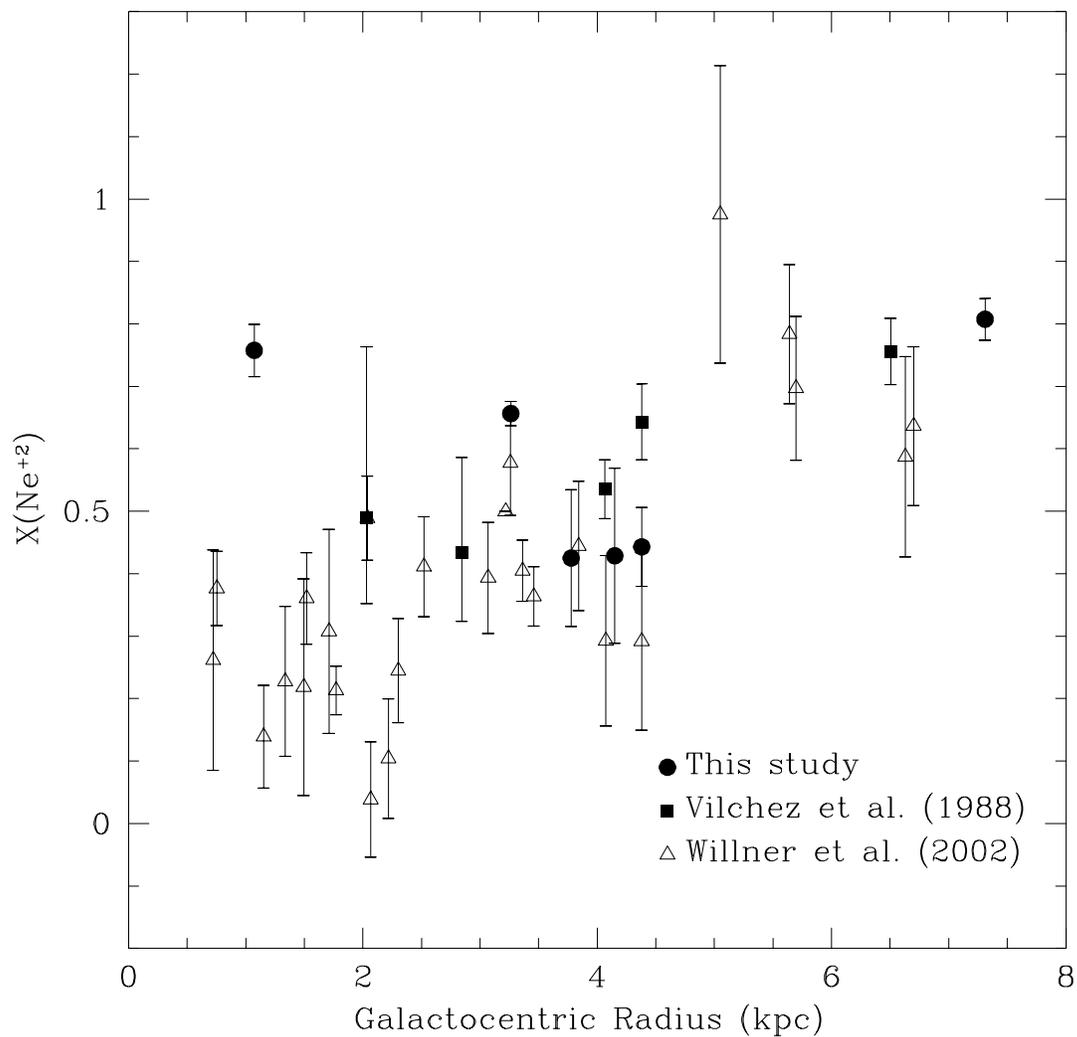}
\caption[ne_ionize.ps]{Comparison of neon ionization derived from data in
this study and V\'\i lchez et al. (1988) to those derived in the Willner
study as a function of galactocentric radius. Symbols are the same as in
Figure 6. No offset in neon ionization is apparent except for BCLMP094
which is a He~II nebula.}
\end{figure}

\clearpage

\begin{figure}
\plotone{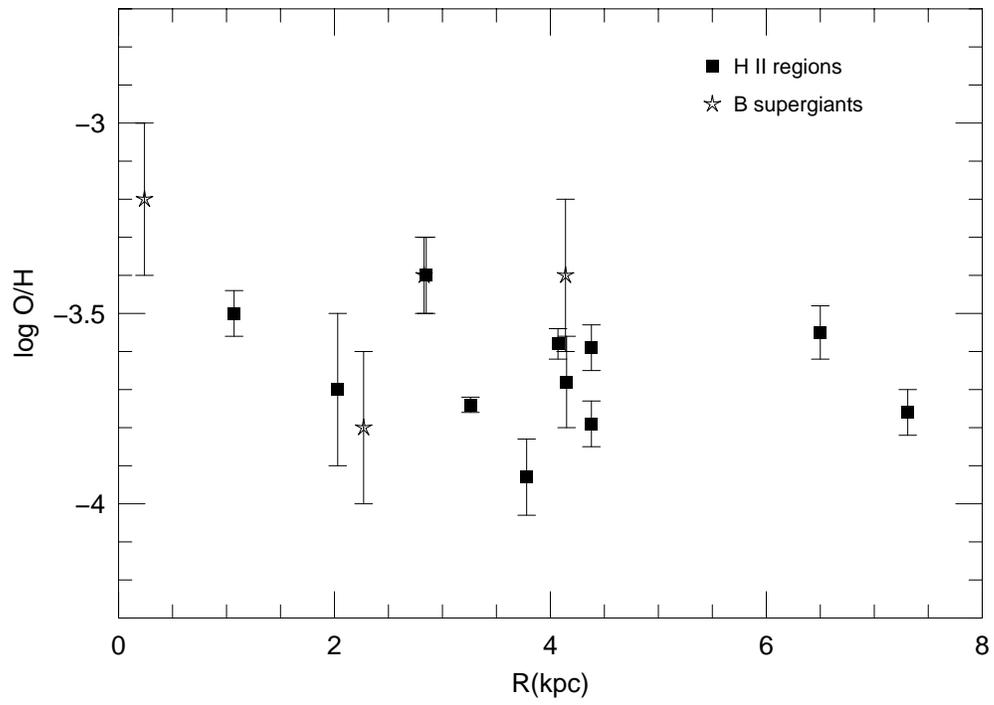}
\caption[fig8.ps]{Comparison of oxygen abundances in M33 \hii\ regions
($filled~squares$ -- this work) and B supergiants ($stars$ -- Monteverde
et al. 2000).}
\end{figure}

\clearpage

\begin{figure}
\plotone{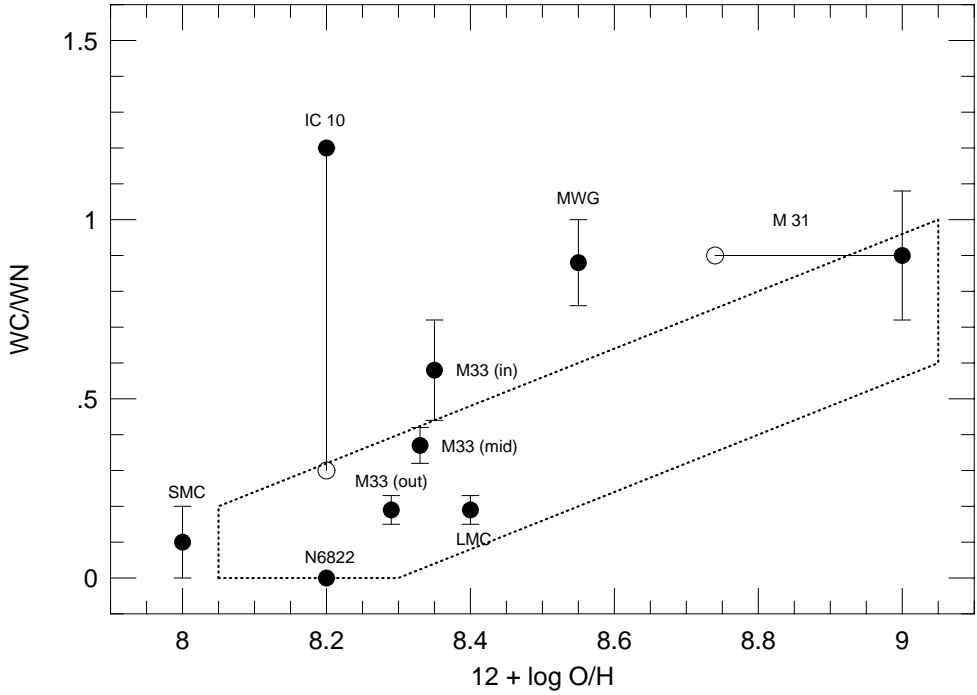}
\caption[fig9.ps]{The observed correlation between the WC/WN number
ratio and oxygen abundance for Local Group galaxies, adapted from Massey
(2003). See that paper for a complete discussion of the Wolf-Rayet star
data. The points for M33 represent rough average values for O/H at 
points 0.8 kpc (in), 2 kpc (mid), and 5 kpc (out) from the nucleus.
The dotted line shows the range of values predicted by Meynet \&
Maeder (2005) for rotating and non-rotating stars.}
\end{figure}

\clearpage

\begin{deluxetable}{lcccc}
\tablecolumns{5}
\tablewidth{0pt}
\tabletypesize{\footnotesize}
\tablecaption{\hii\ Region Coordinates And Derived Galactocentric Radii}

\tablehead{\colhead{ID\tablenotemark{a}} & \colhead{R.A.} & \colhead{Decl.} & \colhead{P.A.} & \colhead{Rad.} \\
\colhead{} & \colhead{(J2000.0)} & \colhead{(J2000.0)} & \colhead{deg.} & \colhead{pc}}

\startdata
BCLMP016  & $1:33:50.0$ & $30:37:32.4$ & $53.7$  & $552$  \\
BCLMP027  & $1:33:45.9$ & $30:36:50.4$ & $53.7$  & $720$  \\
BCLMP094  & $1:34:00.4$ & $30:38:07.6$ & $90.4$  & $769$  \\
BCLMP090  & $1:34:04.2$ & $30:38:09.2$ & $90.4$  & $1073$ \\
BCLMP696  & $1:33:58.4$ & $30:48:40.1$ & $157.2$ & $2345$ \\
CPDSP212  & $1:33:57.7$ & $30:49:02.2$ & $157.2$ & $2456$ \\
BCLMP691  & $1:34:16.6$ & $30:51:54.0$ & $90.0$  & $3263$ \\
BCLMP745  & $1:34:37.6$ & $30:34:55.0$ & $78.1$  & $3778$ \\
BCLMP705  & $1:34:40.3$ & $30:31:20.5$ & $53.3$  & $4017$ \\
BCLMP740  & $1:34:39.5$ & $30:41:47.9$ & $90.4$  & $4071$ \\
BCLMP706  & $1:34:42.2$ & $30:31:42.3$ & $53.3$  & $4150$ \\
BCLMP290A & $1:33:11.4$ & $30:45:15.1$ & $112.0$ & $4380$ \\
MA1\tablenotemark{b}       & $1:33:03.4$ & $30:11:18.7$ & $140.4$ & $7310$ \\ 
\enddata

\tablenotetext{a}{A BCLMP label denotes an ID from Boulesteix et al. 1974. A CPDSP denotes
a label from Court\`es et al. 1987, an extention of the Boulesteix study.} 
\tablenotetext{b}{MA1 does not have a BCLMP or CPDSP ID because it is positioned too
far South for both surveys.}
\end{deluxetable}

\clearpage

\begin{deluxetable}{lccccccc}
\tablecolumns{8}
\tablewidth{0pt}
\tabletypesize{\scriptsize}
\tablenum{2}
\tablecaption{Dereddened Line Fluxes with Uncertainties}
\tablehead{
\colhead{Wavelength} & \colhead{BCLMP016} & \colhead{BCLMP027} & \colhead{BCLMP094}  & \colhead{BCLMP090} & 
\colhead{BCLMP696} & \colhead{CPDSP212} & \colhead{BCLMP691}} 

\startdata
3725 [O~II]      &0.83 $\pm$ 0.08   &1.11 $\pm$ 0.08   &0.62 $\pm$ 0.06   &0.80 $\pm$ 0.10   &1.12 $\pm$ 0.10   &1.15 $\pm$ 0.11   &0.73 $\pm$ 0.04  \\  
3727 [O~II]      &1.23 $\pm$ 0.10   &1.59 $\pm$ 0.11   &0.89 $\pm$ 0.07   &1.18 $\pm$ 0.13   &1.67 $\pm$ 0.13   &1.73 $\pm$ 0.15   &0.95 $\pm$ 0.05    \\ 
3750 H12        &0.04 $\pm$ 0.02   &0.04 $\pm$ 0.02   &0.03 $\pm$ 0.02   &$<$0.022          &$<$0.027          &$<$0.015          &0.032$\pm$ 0.004  \\
3771 H11        &0.04 $\pm$ 0.02   &0.05 $\pm$ 0.01   &0.02 $\pm$ 0.01   &$<$0.018         &$<$0.025           &0.03 $\pm$ 0.01   &0.041$\pm$ 0.004  \\
3798 H10        &0.05 $\pm$ 0.02   &0.06 $\pm$ 0.01   &0.05 $\pm$ 0.01   &0.07 $\pm$ 0.02   &0.06 $\pm$ 0.02   &0.06 $\pm$ 0.02   &0.056$\pm$ 0.004  \\
3835 H9         &0.07 $\pm$ 0.02   &0.07 $\pm$ 0.01   &0.07 $\pm$ 0.01   &0.07 $\pm$ 0.01   &0.07 $\pm$ 0.02   &0.07 $\pm$ 0.01   &0.074$\pm$ 0.004  \\
3865 [Ne~III]    &$<$0.012          &0.03 $\pm$ 0.01   &0.054 $\pm$ 0.008 &0.60 $\pm$ 0.05   &$<$0.011          &0.029 $\pm$ 0.008 &0.25$\pm$0.01    \\
3889 H8+He~I     &0.17 $\pm$ 0.02   &0.17 $\pm$ 0.02   &0.20 $\pm$ 0.01   &0.20 $\pm$ 0.02   &0.14 $\pm$ 0.01   &0.18 $\pm$ 0.02   &0.201$\pm$ 0.008  \\
4026 He~I        &$<$0.008          &0.010 $\pm$ 0.006 &0.016 $\pm$ 0.005 &0.018 $\pm$ 0.007 &$<$0.009          &0.016 $\pm$ 0.006 &0.020 $\pm$ 0.002  \\
4068 [S~II]      &$<$0.009          &0.015 $\pm$ 0.007 &0.010 $\pm$ 0.004 &0.027 $\pm$ 0.008 &0.020 $\pm$ 0.008 &0.018 $\pm$ 0.005 &0.015 $\pm$ 0.001  \\
4100 H$\delta$  &0.26 $\pm$ 0.02   &0.26 $\pm$ 0.02   &0.26 $\pm$ 0.01   &0.26 $\pm$ 0.02   &0.26 $\pm$ 0.02   &0.26 $\pm$ 0.02   &0.257 $\pm$ 0.009  \\
4338 H$\gamma$  &0.47 $\pm$ 0.02   &0.47 $\pm$ 0.02   &0.47 $\pm$ 0.02   &0.47 $\pm$ 0.03   &0.47 $\pm$ 0.02   &0.47 $\pm$ 0.03   &0.47 $\pm$0.02    \\
4363 [O~III]     & $<$ 0.005 &$<$0.005          &$<$0.004          &0.041 $\pm$ 0.006 &$<$0.007          &$<$0.004          &0.022 $\pm$ 0.001  \\
4471 He~I        &0.036 $\pm$ 0.006 &0.035 $\pm$ 0.007 &0.041 $\pm$ 0.005 &0.047 $\pm$ 0.007 &0.011 $\pm$ 0.006 &0.031 $\pm$ 0.005 &0.043 $\pm$ 0.002  \\
4686 He~II	&$<$0.006   &$<$0.007   &$<$0.004   &0.039$\pm$0.007  &$<$0.005  &$<$0.004  &$<$0.001  \\
4859 H$\beta$   &1.00 $\pm$ 0.05   &1.00 $\pm$ 0.04   &1.00 $\pm$ 0.04   &1.00 $\pm$ 0.05   &1.00 $\pm$ 0.04   &1.00 $\pm$ 0.05   &1.00 $\pm$0.03  \\  
4959 [O~III]     &0.20 $\pm$ 0.02   &0.27 $\pm$ 0.02   &0.39 $\pm$ 0.02   &2.16 $\pm$ 0.10   &0.11 $\pm$ 0.01   &0.28 $\pm$ 0.02   &1.16$\pm$0.04    \\
5007 [O~III]     &0.59 $\pm$ 0.04   &0.78 $\pm$ 0.04   &1.17 $\pm$ 0.06   &6.60 $\pm$ 0.30   &0.33 $\pm$ 0.02   &0.83 $\pm$ 0.05   &3.40$\pm$0.12    \\
$C_{H\beta}$    &0.12 $\pm$ 0.09   &0.08 $\pm$ 0.08   &0.32 $\pm$ 0.07  &0.07 $\pm$ 0.10   &0.41 $\pm$ 0.09   &0.21 $\pm$ 0.09   &0.01$\pm$0.04    \\
EW (A)          &0.6               &0.9               &0.2               &2.7               &0.7               &1.0                &0.6     \\
\enddata
\end{deluxetable}


\begin{deluxetable}{lcccccc}
\tablecolumns{7}
\tablewidth{0pt}
\tabletypesize{\scriptsize}
\tablenum{3}
\tablecaption{Dereddened Line Fluxes with Uncertainties}
\tablehead{
\colhead{Wavelength} & \colhead{BCLMP745} & \colhead{BCLMP705}   & \colhead{BCLMP740} & \colhead{BCLMP706} & 
\colhead{BCLMP290A} & \colhead{MA1}}

\startdata
3725 [O~II]      &0.87 $\pm$ 0.07    &1.33 $\pm$ 0.11    &1.22 $\pm$ 0.08    &1.09 $\pm$ 0.12    &0.89 $\pm$ 0.04    &0.46 $\pm$ 0.03 \\  
3727 [O~II]      &1.26 $\pm$ 0.10    &2.06 $\pm$ 0.17    &1.73 $\pm$ 0.11    &1.70 $\pm$ 0.17    &1.34 $\pm$ 0.06    &0.68 $\pm$ 0.05  \\  
3750 H12        &0.04 $\pm$ 0.02    &0.06 $\pm$ 0.01    &0.03 $\pm$ 0.01    &0.05 $\pm$ 0.02    &0.042 $\pm$ 0.004  &$<$0.016  \\          
3771 H11        &0.05 $\pm$ 0.02    &0.07 $\pm$ 0.01    &0.05 $\pm$ 0.01    &0.06 $\pm$ 0.02    &0.046 $\pm$ 0.003  &$<$0.014   \\       
3798 H10        &0.05 $\pm$ 0.01    &0.07 $\pm$ 0.01    &0.06 $\pm$ 0.02    &0.05 $\pm$ 0.02    &0.057 $\pm$ 0.005  &$<$0.016  \\         
3835 H9         &0.07 $\pm$ 0.01    &0.07 $\pm$ 0.02    &0.074 $\pm$ 0.009  &0.07 $\pm$ 0.02    &0.074 $\pm$ 0.005  &0.07 $\pm$ 0.01  \\  
3865 [Ne~III]    &0.08 $\pm$ 0.01    &$<$0.013           &0.056 $\pm$ 0.006  &0.14 $\pm$ 0.02    &0.084 $\pm$ 0.005  &0.41 $\pm$ 0.02  \\ 
3889 H8+He~I     &0.16 $\pm$ 0.02    &0.17 $\pm$ 0.02    &0.18 $\pm$ 0.01    &0.21 $\pm$ 0.02    &0.186 $\pm$ 0.008  &0.16 $\pm$ 0.01 \\   
4026 He~I        &0.014 $\pm$ 0.005  &$<$0.007           &0.010 $\pm$ 0.004  &0.022 $\pm$ 0.009  &0.011 $\pm$ 0.002  &$<$0.009    \\     
4068 [S~II]      &0.020 $\pm$ 0.005  &0.021 $\pm$ 0.007  &0.022 $\pm$ 0.004  &0.017 $\pm$ 0.008  &0.008 $\pm$ 0.002  &$<$0.007   \\       
4100 H$\delta$  &0.26 $\pm$ 0.02    &0.26 $\pm$ 0.02    &0.26 $\pm$ 0.01    &0.26 $\pm$ 0.02    &0.259 $\pm$ 0.009  &0.25 $\pm$ 0.02 \\   
4338 H$\gamma$  &0.47 $\pm$ 0.03    &0.47 $\pm$ 0.03    &0.47 $\pm$ 0.02    &0.47 $\pm$ 0.03    &0.47 $\pm$ 0.02    &0.48 $\pm$ 0.02 \\   
4363 [O~III]     &0.013 $\pm$ 0.004  &$<$0.007           &$<$0.004           &0.012 $\pm$ 0.005  &0.011 $\pm$ 0.002  &0.043 $\pm$ 0.006 \\
4471 He~I        &0.034 $\pm$ 0.005  &0.020 $\pm$ 0.007  &0.035 $\pm$ 0.004  &0.040 $\pm$ 0.007  &0.035 $\pm$ 0.002  &0.016 $\pm$ 0.006 \\
4686 He~II       &$<$0.003           &$<$0.005           &$<$0.003           &$<$0.006           &$<$0.001           &0.049$\pm$0.004 \\
4859 H$\beta$   &1.00 $\pm$ 0.04    &1.00 $\pm$ 0.05    &1.00 $\pm$ 0.04    &1.00 $\pm$ 0.05    &1.00 $\pm$ 0.03    &1.00 $\pm$ 0.03  \\
4959 [O~III]     &0.58 $\pm$ 0.03    &0.19 $\pm$ 0.02    &0.44 $\pm$ 0.03    &0.72 $\pm$ 0.07    &0.61 $\pm$ 0.02    &1.68 $\pm$ 0.06  \\ 
5007 [O~III]     &1.70 $\pm$ 0.08    &0.59 $\pm$ 0.05    &1.30 $\pm$ 0.07    &2.19 $\pm$ 0.14    &1.88 $\pm$ 0.07    &5.30 $\pm$ 0.20  \\  
$C_{H\beta}$    &0.29 $\pm$ 0.09    &-0.17 $\pm$ 0.10   &0.13 $\pm$ 0.07    &0.22 $\pm$ 0.11    &-0.10 $\pm$ 0.05   &0.01 $\pm$ 0.07  \\ \hline
EW (A)          &1.5                &1.8                &0.6                &1.2                &0.7                &0.6 \\               
\enddata
\end{deluxetable}

\clearpage

\begin{deluxetable}{lcc}
\tablecolumns{3}
\tablewidth{0pt}
\tabletypesize{\small}
\tablenum{4}
\tablecaption{Measured Electron Temperatures and Densities}
\tablehead{\colhead{ID} & \colhead{T[O~III]} & \colhead{$n_{e}$} \\ 
           \colhead{}   & \colhead{(K)}         & \colhead{(cm$^{-3}$)}}

\startdata
BCLMP016  & ...                       & $<180$ \\
BCLMP027  & ...                       & $<170$ \\
BCLMP094  & ...                       & $<210$ \\
BCLMP090  & $9,900 \pm^{500}_{400}$   & $<250$ \\
BCLMP696  & ...                       & $<160$ \\
CPDSP212  & ...                       & $<170$ \\
BCLMP691  & $10,000 \pm^{200}_{200}$  & $100\pm60$  \\
BCLMP745  & $10,600 \pm^{1200}_{600}$ & $<190$ \\
BCLMP705  & ...                       & $<130$ \\
BCLMP740  & ...                       & $<170$ \\
BCLMP706  & $9,600 \pm^{1300}_{600}$  & $<170$ \\
BCLMP290A & $9,800 \pm^{600}_{400}$   & $<80$  \\
MA1       & $10,800 \pm^{600}_{400}$  & $<150$ \\
\enddata
\end{deluxetable}

\clearpage

\begin{deluxetable}{lccccc}
\tablecolumns{6}
\tablewidth{0pt}
\tabletypesize{\small}
\tablenum{5}
\tablecaption{Ionic and Elemental Abundances}
\tablehead{\colhead{ID} & \colhead{$O^{+}/H^{+}(\times10^{5})$} & \colhead{$O^{+2}/H^{+}(\times10^{5})$ } & \colhead{$Ne^{+2}/H^{+}(\times10^{5})$} & 
\colhead{log(O/H)} & \colhead{log(Ne/H)}} 

\startdata
BCLMP090  & $7.6  \pm 1.2$           & $23.8 \pm 3.9$           & $6.8 \pm 1.4$         &  $-3.50 \pm 0.06$           &  $-4.04 \pm 0.09$          \\ 
BCLMP691  & $6.3  \pm 0.4$           & $12.0 \pm 0.7$           & $2.7 \pm 0.2$         &  $-3.74 \pm 0.02$           &  $-4.38 \pm 0.03$          \\
BCLMP745  & $6.8  \pm 2.0$           & $5.0  \pm 1.7$           & $0.7 \pm 0.3$         &  $-3.93 \pm 0.10$           &  $-4.8  \pm 0.2$           \\
BCLMP706  & $12.0 \pm 4.5$           & $9.0  \pm 3.9$           & $1.9 \pm 1.0$         &  $-3.68 \pm 0.12$           &  $-4.4  \pm 0.3$           \\
BCLMP290A & $9.0  \pm 1.5$           & $7.2  \pm 1.4$           & $1.0 \pm 0.2$         &  $-3.79 \pm 0.06$           &  $-4.6  \pm 0.1$           \\
MA1       & $3.4  \pm 0.5$           & $14.2 \pm 2.2$           & $3.3 \pm 0.6$         &  $-3.76 \pm 0.06$           &  $-4.39 \pm 0.08$          \\
\enddata
\end{deluxetable}

\clearpage

\begin{deluxetable}{lcccc}
\tablecolumns{5}
\tablewidth{0pt}
\tablenum{6}
\tablecaption{Derived Galactocentric Raddi and Electron Temperatures for  \citet{vilc88} \hii\ Regions}
\tablehead{\colhead{ID} & \colhead{Rad.}  & \colhead{$T[S~III]$} & \colhead{$T[O~II]$} & \colhead{$T[O~III]$} \\
           \colhead{}   & \colhead{(pc)}  & \colhead{(K)}       & \colhead{(K)}      & \colhead{(K)} }

\startdata
MA2      & $2033$  & ...                      & $9,000 \pm^{1300}_{700}$ & ...                      \\
NGC595   & $2846$  & $7,700 \pm^{600}_{400}$  & ...                      & ...                      \\  
NGC604   & $4066$  & ...                      & $9,100 \pm^{300}_{200}$  & $8,500 \pm^{300}_{200}$  \\
IC131 (BCLMP290A)    & $4380$  & ...                      & ...                      & $9,500 \pm^{600}_{400}$  \\
NGC588   & $6505$  & ...                      & ...                      & $9,300 \pm^{600}_{400}$  \\
\enddata
\end{deluxetable}

\clearpage

\begin{deluxetable}{lccccc}
\tablecolumns{6}
\tablewidth{0pt}
\tabletypesize{\small}
\tablenum{7}
\tablecaption{Ionic and Elemental Abundances for \citet{vilc88} \hii\ Regions}
\tablehead{\colhead{ID} & \colhead{$O^{+}/H^{+}(\times10^{5})$} & \colhead{$O^{+2}/H^{+}(\times10^{5})$ } & \colhead{$Ne^{+2}/H^{+}(\times10^{5})$} & 
\colhead{log(O/H)} & \colhead{log(Ne/H)}} 

\startdata
MA2       & $11.2 \pm^{7.7}_{4.6}$   & $10.8 \pm^{9.2}_{4.0}$   & $2.0 \pm^{2.1}_{0.7}$ &  $-3.7  \pm^{0.2}_{0.1}$    &  $-4.4  \pm^{0.5}_{0.2}$   \\
NGC595    & $23.3 \pm^{10.6}_{7.8}$  & $17.8 \pm^{7.5}_{5.3}$   & ...                   &  $-3.4  \pm 0.1$            &  ...                       \\
NGC604    & $12.1 \pm 1.6$           & $14.0 \pm 1.9$           & $2.5 \pm 0.4$         &  $-3.58 \pm 0.04$           &  $-4.33 \pm 0.08$          \\
IC131 (BCLMP290A) & $9.2  \pm 1.6$           & $16.6 \pm 3.3$           & $6.5 \pm 1.9$         &  $-3.59 \pm 0.06$           &  $-4.0  \pm 0.1$           \\
NGC588    & $6.8  \pm 1.3$           & $21.2 \pm 4.6$           & $6.1 \pm 1.6$         &  $-3.55 \pm 0.07$           &  $-4.1  \pm 0.1$           \\
\enddata
\end{deluxetable}

\end{document}